\documentclass[]{aa}
\usepackage{txfonts}
\usepackage{natbib}
\usepackage{graphicx}
\usepackage{psfig}
\usepackage{amssymb}

\def\src{XTE~J1739-302}

\def\xmm{{\em XMM--Newton}}

\def\inte{{\em INTEGRAL}}
\def\suz{{\em Suzaku}}
\def\chandra{{\em Chandra}}
\def\nustar{{\em NuSTAR}}
\def\xte{{\em RossiXTE}}

\def\gaia{{\em Gaia}}
\def\epic{{EPIC}}

\def\approxgt{\mathrel{\hbox{\rlap{\lower.55ex \hbox {$\sim$}}
        \kern-.3em \raise.4ex \hbox{$>$}}}}
\def\approxlt{\mathrel{\hbox{\rlap{\lower.55ex \hbox {$\sim$}}
        \kern-.3em \raise.4ex \hbox{$<$}}}}

\def\flux {\mbox{erg cm$^{-2}$ s$^{-1}$}}
\def\lum {\mbox{erg s$^{-1}$}}
\def\nh{$N_{\rm H}$}
\def\ltsima{$\; \buildrel < \over \sim \;$}
\def\lsim{\lower.5ex\hbox{\ltsima}}
\def\gtsima{$\; \buildrel > \over \sim \;$}
\def\gsim{\lower.5ex\hbox{\gtsima}}

\def\hcm {\hbox {\ifmmode $ atom cm$^{-2}\else atom cm$^{-2}$\fi}}
\def\arcmin {\hbox{$^\prime$}}
\def\arcsec {\hbox{$^{\prime\prime}$}}

\def \apj {ApJ}
\def \aj {AJ}
\def \apjl {ApJL}
\def \apjs {ApJS}
\def \aap {A\&A}

\def \mnras {MNRAS}

\def \ssr {Space Science Reviews}

\def \nar {New Astronomy Reviews}

\newcommand{\be}{\begin{equation}}

\newcommand{\ee}{\end{equation}}

\newcommand{\swift}{{\emph{Swift}}}

\begin{document}
\title{Capturing the lowest luminosity state of the Supergiant Fast X-ray Transient \src
}

\author{L.~Sidoli\inst{1},   G.~Ponti\inst{2,3}, V.~Sguera\inst{4}, P. Esposito\inst{5,1} 
}
\institute{$^1$ INAF, Istituto di Astrofisica Spaziale e Fisica Cosmica, via A.\ Corti 12, 20133 Milano, Italy
\\
$^2$ INAF, Osservatorio Astronomico di Brera, via E.\ Bianchi 46, I-23807 Merate (LC), Italy \\
$^3$ Max-Planck-Institut f{\"u}r extraterrestrische Physik, Giessenbachstrasse, D-85748, Garching, Germany \\
$^4$ INAF, Osservatorio di Astrofisica e Scienza dello Spazio, Via P.\ Gobetti 101, 40129 Bologna, Italy \\
$^5$ Scuola Universitaria Superiore IUSS Pavia, Piazza della Vittoria 15, 27100, Pavia, Italy 
}

\offprints{L. Sidoli, lara.sidoli@inaf.it}

\date{Received 21 December 2022 / Accepted 24 January 2023}

\authorrunning{L. Sidoli et al.}

\titlerunning{\src\ in low X-ray luminosity state}

\abstract{
  We report here on the results of the analysis of \chandra, \xmm\ and \nustar\ recent observations of the Supergiant Fast X-ray Transient  \src. 
  The source was caught in a low X-ray luminosity state, from a few $10^{31}$ to $10^{34}$ \lum\ (0.5-10 keV).
 In particular, a very low X-ray luminosity was captured during an \xmm\ observation performed in October 2022, at a few $10^{31}$ \lum\ (0.5-10 keV), never observed before in \src.
 The \xmm\ spectrum could be well fitted either by an absorbed, steep power law model (photon index of 3.5) or by a collisionally-ionized diffuse gas with a temperature of 0.7 keV, very likely produced by shocks in the supergiant donor wind.
  These observations covered different orbital phases, but all appear compatible with the low luminosity level expected from the orbital \inte\ light curve. The absorbing column density is variable in the range $10^{22}-10^{23}$\,cm$^{-2}$.
  A broad-band X-ray spectrum could be investigated at $10^{34}$ \lum\ (0.5-30 keV) for the first time in \src\ with not simultaneous (but at similar orbital phases) \chandra\ and \nustar\ data, showing a power law spectral shape with a photon index of $\sim$2.2 and an absorbing column density of $\sim$$10^{23}$\,cm$^{-2}$.
    Remarkably, owing to the \xmm\ observation, the amplitude of the X-ray variability has increased to five orders of magnitude, making \src\ one of the most extreme SFXTs.

}

\maketitle

        \section{Introduction\label{intro}}

The X-ray surveys of the Milky Way performed with focusing instruments are legacy datasets of paramount importance for many kind of investigations, from the study of
diffuse X-ray emitting structures to the faintest point-like sources
(e.g. \citealt{Muno2009, Ponti2015sgra, Ponti2015, Degenaar2010, Degenaar2015, Bodaghee2014}).
One of their important outcomes is the monitoring and characterization of the behaviour of  transient sources at their lowest X-ray luminosity state.
In this paper we make use of observations taken from two ongoing survey programmes of the Galactic Center region performed with \chandra\ and \xmm\ in 2022 to cast light on the behavior outside outbursts of the Supergiant Fast X-ray Transient \src.

Supergiant Fast X--ray Transients (SFXTs) are a subclass of high mass X-ray binaries (HMXBs) recognized through \inte\ observations \citep{Sguera2005, Sguera2006, Negueruela2005}. Their outbursts are brief (lasting less than a few days; \citealt{Romano2007, Rampy2009}) and made of a number of bright X-ray flares with a duration of a few thousand seconds, when the X-ray luminosity reaches $\sim$10$^{36}$~\lum. 
These flares show a power law distribution of their hard X-ray luminosity \citep{Smith2012, Paizis2014} and the outburst duty cycle (i.e the percentage of time spent in bright X-ray flares) is shorter than 5\% \citep{Sidoli2018}.

SFXTs are associated with early type supergiants, as the classical persistent accreting pulsars in HMXBs. But, at odds with persistent HMXBs, they show a large range of X-ray variability between flare peaks and quiescence  
(below 10$^{34}$~\lum) of $\sim10^{2}-10^{4}$ \citep{Sidoli2018}, reaching 10${^6}$ only in the source IGR~J17544-2619 \citep{Romano2015:giant}. 
The physical driver of this phenomenology is very debated in the literature (see \citealt{Kretschmar2019} for the most recent review). 

The source \src\ (also known as IGR\,J17391-3021; \citealt{Sunyaev2003}) was discovered in 1998 \citep{Smith1998} and classified as an SFXT because of its  X-ray emission detected by \inte\  \citep{Sguera2005} and \xte\  \citep{Smith2006} only during short (a few thousand seconds) flares above $\sim$10$^{35}$~\lum\ (for a distance of 2 kpc, see below).
The outburst duty cycle is very low (0.89\%; \citealt{Sidoli2018}), with short bright flares caught by 
$ASCA$ \citep{Sakano2002},
\xte\ \citep{Smith2006, Smith2012}, 
\inte\ \citep{Sunyaev2003, Lutovinov2005, 
Sguera2005, Sguera2006, Walter2007, Blay2008, Drave2010, Ducci2010, Paizis2014, Sidoli2016, Sidoli2018}
and \swift\ \citep{Sidoli2009, Sidoli2009:outburst, 
Romano2009, Farinelli2012, Romano2014, Romano2015, Romano2022}.
The latest bright flaring activity has been detected by \inte/JemX  on October 3, 2021 \citep{Maartensson2021}.

The out-of-outburst emission was investigated by means of $ASCA$ \citep{Sakano2002}, \chandra\ \citep{Smith2006}, \suz\ \citep{Bodaghee2011, Pradhan2021},
\swift\ \citep{Sidoli2008, Romano2015} and \xmm\ \citep{Bozzo2010, Sidoli2019}. 
These observations  caught (or allowed to constrain) the source  X-ray emission in the luminosity range 10$^{32}-10^{34}$~\lum.

An orbital period of 51.47$\pm{0.02}$~days was discovered in \inte\ data, from the modulation of the X--ray light curve \citep{Drave2010}.
The folded \inte\ light curve shows a main peak (at orbital phases 0.4-0.5, assuming a zero phase at epoch MJD 52698.2) 
with two side peaks
covering phases 0.25-0.35 and 0.65-0.75. The shape of the X-ray light curve suggests a large orbital eccentricity \citep{Drave2010}. 
A radial velocity investigation is lacking, therefore the orbital geometry (except for the orbital period) is unknown. 
Nevertheless, the periastron (apastron) passage is usually assumed to be located near the phases of the maximum (minimum) X-ray level of emission in the orbital curve.

The refined \chandra\ position \citep{Smith2004, Smith2006} enabled the optical/near-infrared association 
with an  O8Iab(f) star located at a distance of 2.3$^{+0.6} _{-0.5}$~kpc \citep{Negueruela2006}. 
Within its uncertainties, the nominal value is compatible with the distance derived by \citealt{Rahoui2008} ($\sim$2.7 kpc) from the fitting of the broad-band spectral energy distribution of the companion star.
Precise information on the source distance is mandatory to  calculate the correct associated luminosities. To this aim, nowadays accurate and reliable distance estimates are  available from the \gaia\ mission \citep{gaiamission2016}. We relied on the \gaia\ EDR3 data \citep{gaiacoll2021}, and  the  distances by \citet{bailerjones2021}. 
Two distance estimates are available for \src\ (obtained with two different methods) which are  fully compatible with each other: d$_{geo}$=1.93$^{+0.20}_{-0.16}$ kpc  (based only on the parallax) and d$_{photogeo}$=2.01$^{+0.16}_{-0.16}$ kpc  (which uses also the colour and the apparent magnitude).  
We note that the \gaia\ distance has a smaller uncertainty (1.76-2.13 kpc) compared with the range  derived from optical and infrared observations (1.8-2.9 kpc, \citealt{Negueruela2006}). Therefore, according to the \gaia\ results, we will adopt a distance of 2 kpc in our paper.

\section{Observation and data reduction}
\label{data}
 
We summarize in  Table~\ref{tab:log} the observations investigated here, together with the orbital phases covered assuming the ephemeris of \cite{Drave2010}.
The spectral analysis was performed using {\tt xspec} 
in  {\tt HEASoft} (\citealt{heasoft}; v.29).
We adopted the absorption model {\tt TBabs} to fit the spectrum, assuming the photoelectric absorption cross sections of \cite{Verner1996} and the interstellar abundances of \cite{Wilms2000}. 
The spectra were grouped to have at least 15 counts per bin, to apply the $\chi^{2}$ statistics. In low counts spectra, 1 count per bin was adopted and Cash statistics \citep{Cash1979}.
All  uncertainties  in the spectral analysis are given at 90\% confidence level, for one interesting parameter. 
Uncertainties on the count rates are at 1~$\sigma$. 
The uncertainty on the unabsorbed X-ray fluxes have been calculated using {\tt cflux} in {\tt xspec}.
When spectra from different instruments were  simultaneously fitted in {\tt xspec}, we  adopted numerical factors to account for the calibration uncertainties.
The arrival times of all events were corrected to the Solar System barycenter. 
In the following sub-sections the data reduction and analysis specific to the different instruments are explained.

\subsection{\chandra}
\chandra\ serendipitously observed the source sky position in 2022 during the survey program of
the Galactic Center region (Table\,\ref{tab:log};  PI G.~Ponti; Program IDs GDHIdp40, GDHIdp44 and GDHIdp51).
All observations were performed with ACIS-I in very faint mode (VFAINT). 
The data   were reprocessed with standard procedures using the \chandra\ Interactive Analysis of Observation  
({\tt CIAO} 4.14) and {\tt CALDB} (4.9.8). 
Images and exposure maps were produced using {\tt fluximage} in the energy range 0.5-7 keV. 
We used the tool  {\tt srcflux}  to estimate the radii of
the circular regions enclosing 90\% of the point spread function at 1.0 keV
at the large off-axis angles of the source location.
The source-free backgrounds were extracted using annular regions centered on the \src\ position, adopting an inner and an outer radius of 
two and five times source extraction radius.
We report the results in Sect.~\ref{sect:chandrares}.

\begin{table*}
\caption[]{Summary of the observations analysed in this paper. 
}
\begin{tabular}{llllllll}
\hline
\noalign {\smallskip}
        & Satellite   &  ObsID      & Start Time  (UTC)              & Stop time (UTC) 	        & Exp.    & Off-axis angle    & Orbital \\
        &             &             & (yyyy-mm-dd hh:mm:ss)          &  (yyyy-mm-dd hh:mm:ss)         & (ks)    & (arcmin)    & phase \\      
\hline
\noalign {\smallskip}
1       & \chandra\   &  24108      & 2022-03-04 09:11:25          &   2022-03-04 10:50:27          &  4.4	  & 9.6       & 0.92      \\   
2       &  \chandra\  &  24151      & 2022-06-21 17:21:30          &   2022-06-21 18:46:16          &  4.2 	  & 5.6       & 0.04       \\   
3       &  \chandra\  &  24086      & 2022-09-06 04:03:08          &   2022-09-06 05:38:27          &  4.2	  & 8.9       & 0.53       \\   
4       &  \xmm\      &  0886121201 & 2022-10-06 14:56:37          &   2022-10-06 21:49:18          &  26.3	  & 1.5       & 0.12        \\   
5       &  \nustar\   & 30601023002 & 2021-02-16 20:51:09          &   2021-02-17 20:26:09          &  41.9        & 0.0       & 0.52-0.54  \\ 
\hline
\label{tab:log}
\end{tabular}
\end{table*}

\begin{figure}
\begin{center}
    \includegraphics[width=5.4cm,angle=-90]{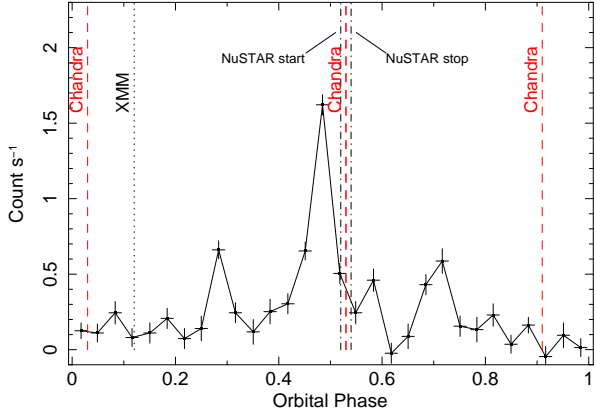}
  \caption{Orbital phases (marked by vertical lines) of the \chandra\ (red dashed lines), \xmm\ (black dotted line) and \nustar\ observations (where we indicate the phase of the start and stop times), overlaid on the \inte/IBIS/ISGRI orbital curve of \src\ we have drawn from \citet{Drave2010}. We remark that the uncertainty on the position of the vertical lines is $\Delta\phi$=$\pm{0.05}$, extrapolating Drave's ephemeris at the time of these observations.
}
\label{fig:orbit}
\end{center}
\end{figure}

\subsection{\xmm}

The source sky position was serendipitously observed with \xmm\ \citep{Jansen2001} in October 2022 during a survey of
the central portion of the Milky Way disc (Table\,\ref{tab:log}; PI G.~Ponti; Program~ID MW~Plane155).
Luckily, \src\ was imaged almost on-axis.
During this pointing  (ObsID. 0886121201)  the three European Photon Imaging Cameras (\epic) \citep{Struder2001, Turner2001}
operated in full frame mode, adopting the medium filter.

EPIC data were reprocessed with {\tt epchain} and {\tt emchain} in the version 20 
of the \xmm\ Science Analysis Software (SAS), adopting standard procedures and the most updated calibration files.
We adopted PATTERN selection from 0 to 4 in the EPIC pn and from 0 to 12 in both MOS.
Response matrices were generated using the SAS tasks {\tt arfgen} and {\tt rmfgen}.

Unfortunately, the observation was severely affected by  high background levels. Adopting the same procedure used by \citet{Ponti2015}, background flares were filtered-out reducing the usable exposure time to about 11 ks for each of the two MOS and to 7.5 ks for the EPIC pn.

\begin{figure}
\begin{center}
\includegraphics[width=8.8cm,angle=0]{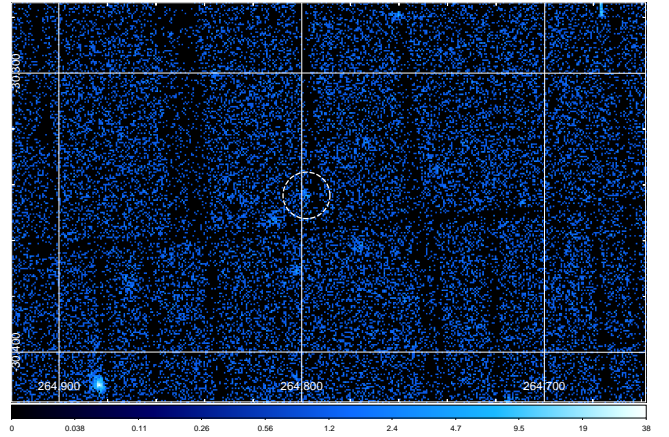}
\caption{Central part of the \xmm\ mosaic of the three EPIC exposures (0.3-12 keV),
  with the faint \src\ marked by the white, dashed circle.
}
\label{fig:xmmmosaic}
\end{center}
\end{figure}

Since a visual inspection of the EPIC mosaic 
revealed a faint source positionally coincident with \src\ (Fig.~\ref{fig:xmmmosaic}), we  performed a proper source detection procedure using the SAS tool {\tt edetect\_chain}. 
It first performs a sliding box source searching (with both local and global background) and then a maximum likelihood multi-source point spread function fitting.
We run  {\tt edetect\_chain} on
the EPIC clean images  extracted in the energy bands 0.2-0.5 keV, 0.5-1.0 keV, 1.0-2.0 keV, 2.0-4.5 keV and 4.5-12 keV,
the standard ones adopted also in the \xmm\ source catalog \citep{Webb2020}.
This led to the detection of a source consistent with the \src\ position (with a probability of 2.9$\times$10$^{-15}$ of detection occurring by chance), considering all EPIC in the full energy range 0.2-12 keV. 
The resulting net count rate is (7.56$\pm{1.51}$)$\times$10$^{-3}$~count~s$^{-1}$ (combining the three EPIC in the 0.2-12 keV energy band). The source is better detected in the energy band 1.0-2.0 keV.

The EPIC spectra  were extracted from the cleaned event files using {\tt xmmselect}, with circular regions with a radius of 30\arcsec.
The background spectra were extracted from larger circular regions, free of sources, from the same CCD.
We report the results in Sect.~\ref{sect:xmmres}.

\subsection{\nustar}

The Nuclear Spectroscopic Telescope Array
(\nustar, \citealt{Harrison2013}) carries two identical co-aligned telescopes
focusing X-ray photons onto two Focal Plane Modules, named A and B (hereafter FPMA and FPMB).
Each FPM contains four solid-state cadmium zinc telluride (CdZnTe) detectors, covering
a 12$\arcmin$$\times$12$\arcmin$ FOV,
providing a spectral resolution of 400~eV (full width at half maximum, FWHM)
at 10 keV and a spatial resolution of 18$''$ (FWHM).

\src\ was observed with \nustar\ in February 2021 (Table\,\ref{tab:log}), collecting net exposure times of 41.9\,ks (FPMA) and 41.5\,ks (FPMB).
We downloaded the public data from HEASARC and performed the reduction with {\tt nupipeline} (v.0.4.9) in 
the \nustar\ data analysis software ($NuSTAR~DAS$), using standard procedures and CALDB version 20220608.

The spectra and the light curves were extracted from the cleaned event files using {\tt nuproducts}. 
Given the faintness of the source, a circle with a  60$''$ radius was used to extract source products, while for the background
an annular region with an inner radius of 100$''$ and an outer radius of 120$''$ was assumed.
Given the stable background, no further filtering was applied.
\nustar\ source light curves and event lists used for the timing analysis were extracted with  {\tt nuproducts} and the keyword ``barycorr=yes'', to correct arrival times to the Solar System barycenter.
The net count rates (3-78 keV) calculated from the whole observation were the following:
0.0150$\pm{0.0011}$ counts~s$^{-1}$  (FPMA) and 
0.0133$\pm{0.0012}$ counts~s$^{-1}$  (FPMB).
We report the results in Sect.~\ref{sect:nures}.

\section{Results}
\label{sect:res}

We report here  the results on the low luminosity state of \src\ as observed with the satellites \chandra, \xmm, and \nustar,
in different epochs and orbital phases.

\subsection{\chandra}
\label{sect:chandrares}

\begin{figure}
\begin{center}
\includegraphics[width=5.8cm,angle=-90]{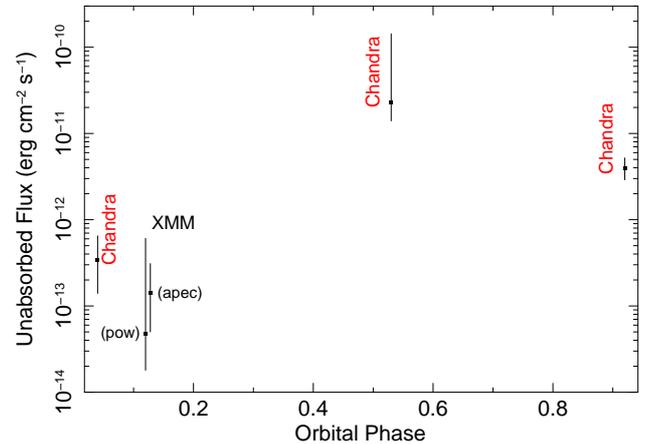} \\
\caption{Source fluxes corrected for the absorption (0.5-10 keV) versus the orbital phase. For the \xmm\ observations we report two values (slightly shifted in orbital phase to better show the error bars) obtained from the two different continuum models adopted, the power law and the {\tt apec} model. We note that for two \chandra\ observations the uncertainties reported in this plot are simply due to the uncertainties in the count rate assuming a fixed spectral shape (see text), while only in the \chandra\ obsID 24086 (at $\phi=0.53$) the flux and its uncertainty are the outcome of the spectral analysis.
}
\label{fig:uforb}
\end{center}
\end{figure}

The source has always been detected in the three observations, but with different intensities: 
we have extracted a spectrum only from  obsID~24086, where the source is brighter,
while for the other two we have estimated an average flux using the {\tt CIAO} tool {\tt srcflux}. 
The spectral parameters are listed in Table~\ref{tab:spec}.
Assuming the same spectral shape obtained from the spectroscopy of ObsID~24086
(an absorbed powerlaw model with a photon index $\Gamma=2.3$, and \nh=$10^{23}$~cm$^{-2}$), 
we obtained the following values:
in ObsID~24108 the net count rate was (9.8$\pm{1.6}$)$\times10^{-3}$~count~s$^{-1}$ (0.5-7 keV),
implying fluxes corrected for the absorption of 3.6($^{+1.1} _{-0.9}$)$\times$10$^{-12}$~\flux\ (0.5-7 keV) 
and 3.9$\times$10$^{-12}$~\flux\ (0.5-10 keV).
This latter value translates into an X-ray luminosity of L$_{0.5-10 keV}$=1.9$\times10^{33}$~\lum.  
In ObsID~24151 the source net count rate was (1.6$\pm{0.6}$)$\times10^{-3}$~count~s$^{-1}$ (0.5-7 keV),
implying fluxes corrected for the absorption of 3.1($^{+2.8} _{-1.8}$)$\times$10$^{-13}$~\flux\ (0.5-7 keV) 
and 3.4$\times$10$^{-13}$~\flux\ (0.5-10 keV).
This latter flux translates into a luminosity of
L$_{0.5-10 keV}$=1.6$\times10^{32}$~\lum.
The fluxes corrected for the absorption (0.5-10 keV) are plotted versus the orbital phase of these observations in Fig.\,\ref{fig:uforb}.
The adoption of the above spectral model is suggested by the faintness of the source: since SFXT spectra are harder when brighter (e.g., \citealt{Romano2014}), the adoption of a harder spectral shape than the one observed during ObsID~24086 (when the source was brighter) would be unusual. We note that \chandra\ observed the source also in 2001 \citep{Smith2006}, with a harder powerlaw spectrum ($\Gamma$=0.62$\pm{0.23}$, \nh=4.2$\pm{1.0}\times10^{22}$~cm$^{-2}$, orbital phase=0.27), but at a much higher flux (1.3$\times$10$^{-11}$~\flux). For what concerns the  column density, the adoption of \nh=$10^{23}$~cm$^{-2}$ should be considered merely as an example, since the absorption can be very variable (also due to the clumpy circumstellar environment). For the sake of completeness, we note that assuming \nh=$10^{22}$~cm$^{-2}$ with the same power law model,  the unabsorbed flux is reduced by a factor of $\sim$4.

\begin{figure*}
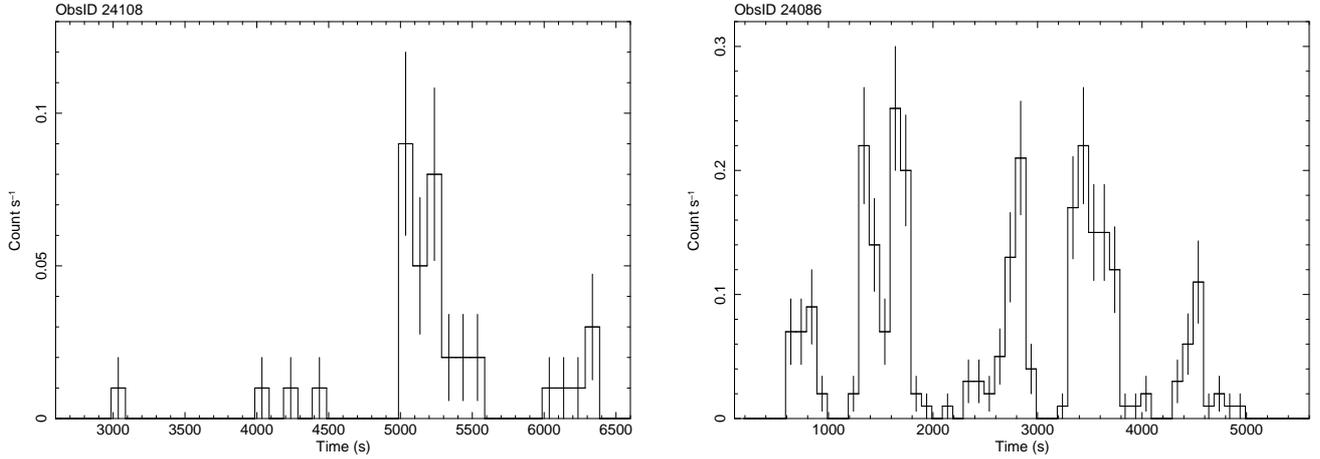

\begin{center}
  \includegraphics[width=6.1cm,angle=-90]{src_24108_sub_lc_100s.ps}
  \includegraphics[width=6.1cm,angle=-90]{src_24086_sub_lc_100s.ps} 
  \caption{\src\ light curves observed by \chandra\ (0.5-7 keV) in March (left panel) and September 2022 (right panel).
}
\label{fig:chalc}
\end{center}
\end{figure*}

The source light curves (0.5-7 keV) during observations  24108 (March 2022) and 24086 (September 2022) are reported in Fig.~\ref{fig:chalc}. We do not show any light curve from observation 24151, because of the limited number of counts collected.

The timing analysis on the barycentered events files from ObsID 24086 yielded a candidate period at $\sim$10.58\,s that, however, after taking into account the number of trials in the Fourier analysis (2048), is not statistically significant ($<$3$\sigma$). The upper limit for the pulsed fraction of a coherent sinusoidal modulation that we derived for the timing series of observation 24086 with extensive simulations is 56\%, in the 0.5--7\,keV band and after the background subtraction. The other \chandra\ observations do not have enough counts for a sensitive timing analysis or to derive meaningful limits.

\subsection{\xmm}
\label{sect:xmmres}

\begin{figure}
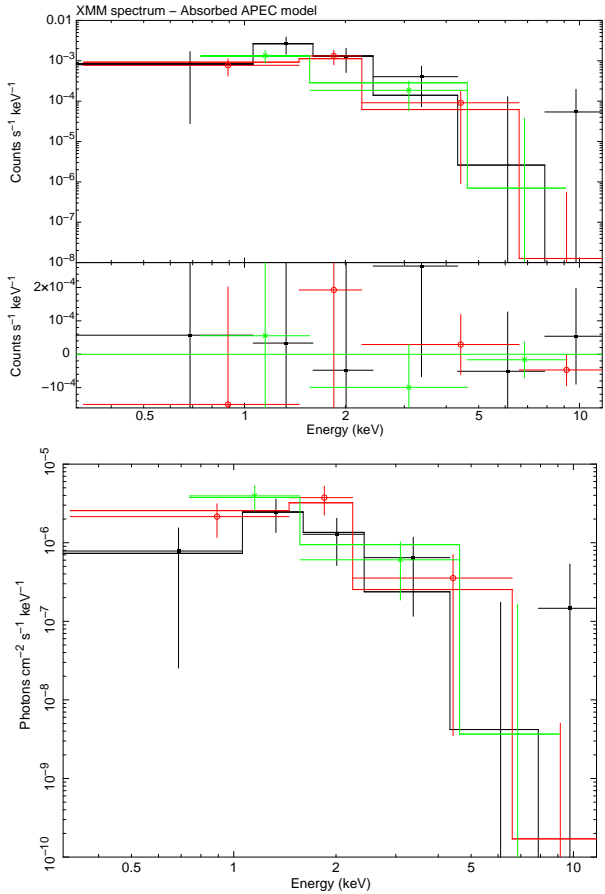

\begin{center}
  \includegraphics[width=5.8cm,angle=-90]{lda_res_apec_xmm_ls.ps} \\
  \includegraphics[width=6.0cm,angle=-90]{ufs_apec_xmm_ls.ps}  
  \caption{\xmm\ spectrum fitted with the {\tt apec} model. The upper panel shows the data together with the residuals. The lower panel displays the photon spectrum. Symbols have the following meaning: EPIC pn is marked by black solid squares, MOS~1 by red open circles and MOS~2 by green crosses. The spectrum has been graphically rebinned, for presentation purposes only.
}
\label{fig:xmmspec}
\end{center}
\end{figure}

We fitted the three EPIC spectra using either an absorbed power law model for the continuum, or an absorbed {\tt apec} model, both resulting in equally good fits. 
The adoption of the {\tt apec} model (an emission spectrum from collisionally-ionized diffuse gas) is suggested by the very low luminosity of the source (a few 10$^{31}$~\lum). 
At these X-ray luminosities, the supergiant strong wind is expected to contribute a significant fraction (if not all) of the X-ray emission (see Sect.\ref{sect:discussion}).
The results are reported in Table\,\ref{tab:spec} and the fit with the {\tt apec} model is shown in Fig.\,\ref{fig:xmmspec}. We searched the barycentered data for periodic modulations, but we found none. The low signal-to-noise ratio of the data did not allow us to derive constraining limits on possible signals.

\subsection{\nustar}
\label{sect:nures}

\begin{figure}
\begin{center}
\includegraphics[width=6.0cm,angle=-90]{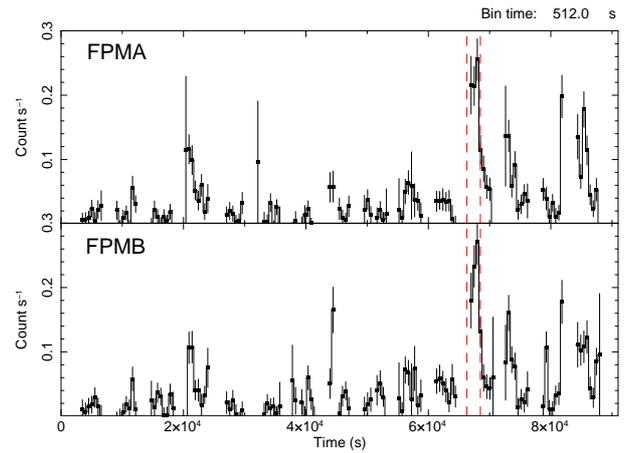}
\caption{\src\ net light curves observed with \nustar\ (3-78 keV). 
The two dashed vertical lines enclose the flare analysed in Sects.~\ref{sect:nures} and \ref{sect:broadspec}.
}
\label{fig:nubaryclc}
\end{center}
\end{figure}

The \nustar\ light curves extracted from both FPMs 
are shown in Fig.\ref{fig:nubaryclc}, where some flaring activity is evident, although the source was not in outburst. 
The observation spanned a whole day, covering the orbital phase
range $\phi$=0.52-0.54. The uncertainty on the orbital phases, extrapolated at the times of these observations, is $\Delta\phi$=$\pm{0.05}$. 

\begin{figure}
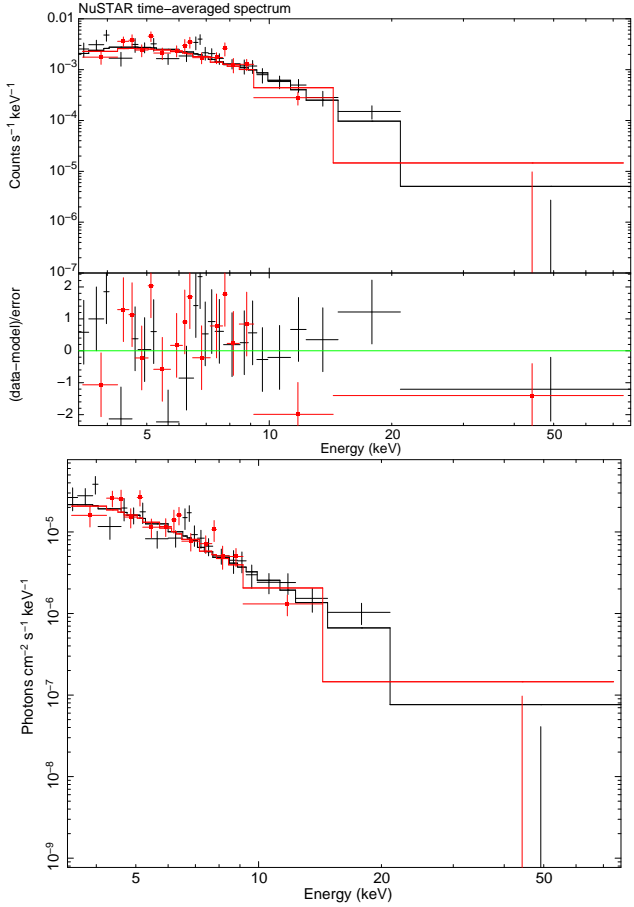

\begin{center}
  \includegraphics[width=6.1cm,angle=-90]{ldadelchi_nu_ab_pow_av_markonB_ed.ps} \\
  \includegraphics[width=5.9cm,angle=-90]{ufs_nu_ab_pow_av_markonB_ed.ps}  
  \caption{\nustar\ spectrum extracted from the whole exposure and fitted with an absorbed power law.
    The upper panel shows the count spectrum together with the residuals in units of standard deviation, while the
    lower panel displays the photon spectrum. Black crosses indicates FPMA, red solid squares marks FPMB data.
}
\label{fig:nuspec}
\end{center}
\end{figure}

\begin{figure}
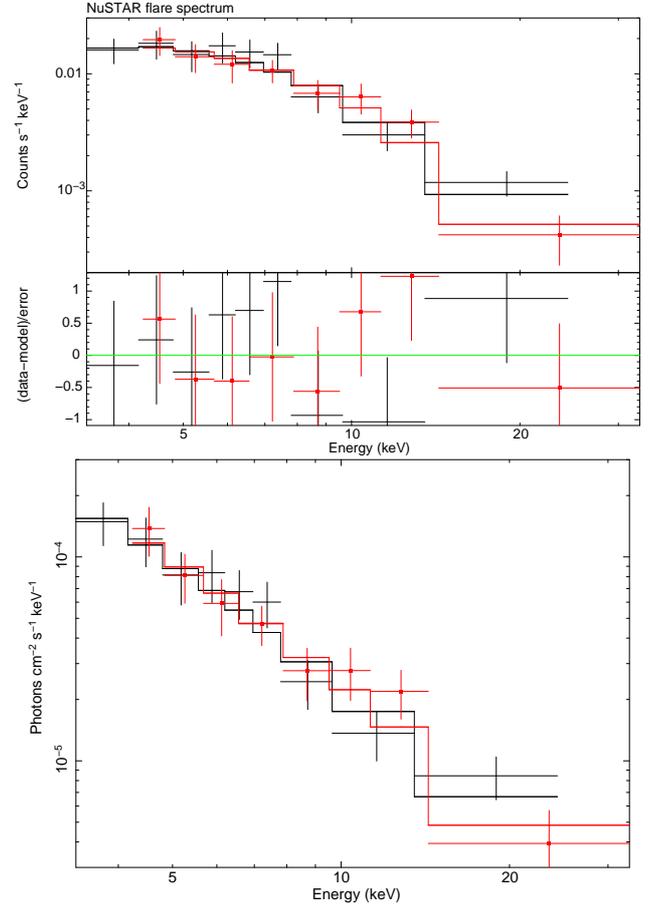

\begin{center}
  \includegraphics[width=6.1cm,angle=-90]{ldadelchi_nu_ab_pow_flare_markonB_ed.ps} \\
  \includegraphics[width=5.9cm,angle=-90]{ufs_nu_ab_pow_flare_markonB_ed.ps}  
  \caption{\nustar\ spectrum extracted from the brightest flare (net exposure time of $\sim$1.5 ks) and fitted with an absorbed power law. Symbols have the same meaning as in Fig.\ref{fig:nuspec}.
}
\label{fig:nuspecflare}
\end{center}
\end{figure}

We have performed two spectral extractions: one from the whole exposure time (time-averaged spectrum in Table~\ref{tab:spec}) and one from the brightest flare present in the light curve (it is marked by vertical dashed lines in Fig.~\ref{fig:nubaryclc}).
A simple absorbed power law is already a good fit to the data, therefore no more complicated models were tried. Spectral parameters are reported in  Table~\ref{tab:spec} for both spectral extractions, while in Fig.~\ref{fig:nuspec} and Fig.~\ref{fig:nuspecflare} we show the time averaged and the flare spectra, respectively.
In Table~\ref{tab:spec},  fourth column, we report also the flux corrected for the absorption extrapolated to a softer energy range, only to compare better with \xmm\ and \chandra\ results. Therefore, these fluxes have no associated uncertainty. However, we note that these values should be taken with caution, as the absorbing column density obtained from \nustar\ spectroscopy only might be unreliable, given the hard energy band of operation (see, e.g., \citealt{Sidoli2017}). 
 
\begin{table*}
  \caption{Spectroscopy of \src. All fluxes are corrected for the absorption. 
}
\label{tab:spec}
\vspace{0.0 cm}
\begin{center}
\begin{tabular}{lllllll} \hline
 \hline
\noalign {\smallskip}
Obs.               &    N$_{\rm H}$            &  Powerlaw $\Gamma$ &  UF$_{\rm 0.5-10 keV}$     & UF$_{\rm 3-78 keV}$ & L$_{\rm 0.5-10 keV}$ &    $\chi^{2}$ (dof)   \\
                   &   (10$^{22}$ cm$^{-2}$)   &   or kT$_{\rm apec}$ (keV) &  (\flux) &  (\flux)        &     (\lum)      &       or C-Stat       \\
\hline
\chandra\ (24086) &  $10^{+6} _{-5}$    & $\Gamma$=$2.3^{+1.3} _{-1.2}$                 &  $2.3^{+12} _{-0.9}\times10^{-11}$ &  $-$     &   1.1$\times10^{34}$ & $\chi^{2}$=16.19 (13 dof)  \\  
\xmm              & $1.1^{+1.3} _{-0.8}$       & $\Gamma$=$3.5^{+2.2} _{-1.5}$              & $4.8^{+56.} _{-3.0}\times10^{-14}$  &  $-$      &  2.3$\times10^{31}$  & C-Stat=209.23 (224 dof) \\   
\xmm              & $2.1^{+0.6} _{-0.5}$       & kT$_{apec}$=$0.70^{+0.24} _{-0.18}$              & $1.4^{+1.7} _{-0.9}\times10^{-13}$   &  $-$   &  6.7$\times10^{31}$  & C-Stat=206.46 (224 dof) \\  
\nustar\ (time-av.) & $16^{+12} _{-11}$  & $\Gamma$=2.8$\pm{0.4}$   &   $6.7\times10^{-12}$ &  $1.60^{+0.35} _{-0.25}\times10^{-12}$ & 3.2$\times10^{33}$  & $\chi^{2}$=116.12 (92 dof) \\ 
\nustar\ (flare) & $3^{+18} _{-3}$   &  $\Gamma$=2.1$^{+0.6} _{-0.3}$ &  $1.3\times10^{-11}$  &  $1.18^{+0.31} _{-0.25}\times10^{-11}$ &  6.2$\times10^{33}$ &  $\chi^{2}$=8.24 (13 dof) \\ 
\chandra\ + \nustar\ (flare)  &  $9.9^{+2.3} _{-2.7}$    & $\Gamma$=$2.20^{+0.33} _{-0.31}$   &  $2.1^{+1.1} _{-0.6}\times10^{-11}$ & $1.54^{+0.33} _{-0.29}\times10^{-11}$      &   1.0$\times10^{34}$ & $\chi^{2}$=24.3 (27 dof)  \\
\hline
\hline
\end{tabular}
\end{center}
\end{table*}

We examined the barycentered light curves and event lists to search for timing modulations. No significant periodic or quasi-periodic signal was found. The upper limits on the pulsed fraction of a sinusoidal periodical signal are 47\% below $\sim$500\,s, and 62\% up to $\sim$2000\,s.  For longer periods, the sensitivity is very low, due to the presence of strong non-Poissonian noise, mainly induced by the \nustar\ orbital period and the source variability.

\subsection{The broad-band X-ray spectrum}
\label{sect:broadspec}

The \chandra\ observation ObsID\,24068 falls within the orbital phase interval
covered by the \nustar\ observation.
From the spectroscopy of these single observations (Table\,\ref{tab:spec}),
it is evident that the \nustar\ spectrum extracted from the whole exposure time captured
a fainter X-ray state than the one caught by \chandra.
On the contrary, the \nustar\ spectrum extracted from the faint flare is consistent with
the one observed by \chandra, and they nicely overlap in the energy range 3-7 keV.
Therefore, we will discuss here the \src\ broad band spectrum (0.5-30 keV), although the non simultaneous 
observations suggest to consider the following results with caution.

\begin{figure}
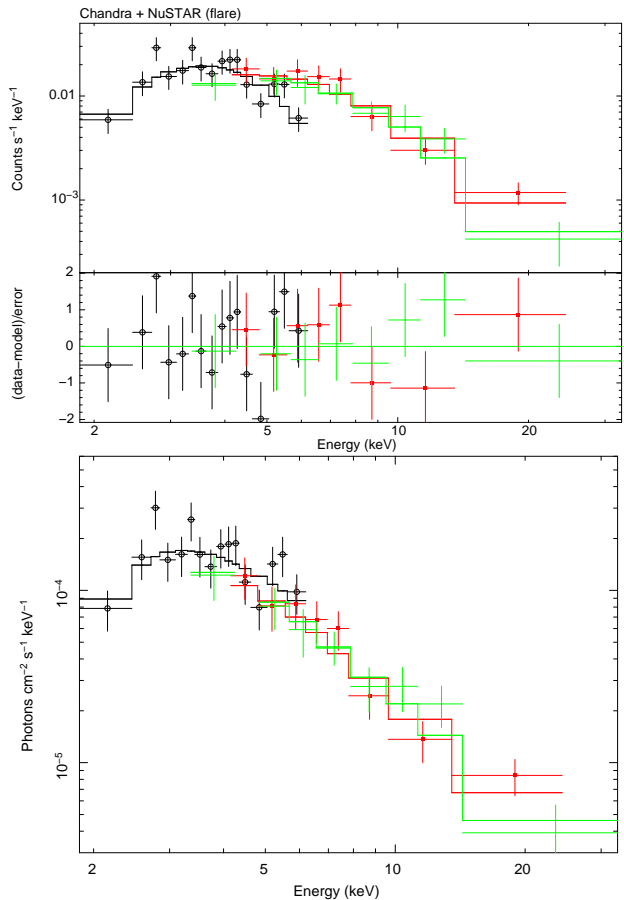

\begin{center}
  \includegraphics[width=6.0cm,angle=-90]{ldadelchi_chandra_nustar_flare_pow_ed_color.ps} \\
  \includegraphics[width=5.9cm,angle=-90]{ufs_chandra_nustar_flare_pow_ed_color.ps}  
  \caption{Broad-band spectrum obtained fitting together \nustar\ spectrum (during the flare) and the \chandra\ observation ObsID\,24086
    (see Sect.\,\ref{sect:broadspec} for details).
    \chandra\ data points are marked with  black open circles, while \nustar\ data are indicated by red solid squares (FPMA) and by green crosses (FPMB). 
}
\label{fig:broadspec}
\end{center}
\end{figure}

Adopting an absorbed power law model we obtained a very good fit ($\chi^{2}$=24.3 with 27 dof), with no evidence
of any curvature in the spectrum or additional spectral components (Fig.\,\ref{fig:broadspec}).
It is important to note that all constant factors adopted to account for calibration uncertainties were consistent with unity:
when the constant factor for \chandra\ is fixed at 1, we got $0.79^{+0.24} _{-0.19}$  and $0.80^{+0.22} _{-0.18}$
for FPMA and FPMB spectra, respectively.

The resulting spectral parameters are an absorbing column density N$_{\rm H}$=$9.9^{+2.3} _{-2.7}$\,$\times10^{22}$\,cm$^{-2}$
and a power law photon index $\Gamma$=$2.20^{+0.33} _{-0.31}$.
We  measured the following fluxes corrected for the absorption:
$2.1 (^{+1.1} _{-0.6}) \times10^{-11}$ \flux\ (0.5-10 keV) and
$3.0 (^{+1.1} _{-0.6}) \times10^{-11}$  \flux, when extrapolated to the 0.5-100 keV energy range.
The X-ray luminosity is 1.4$\times10^{34}$  \lum\ (0.5-100 keV) at 2 kpc.
To the best of our knowledge, this is the first \src\ broad-band spectrum outside outbursts.

\section{Discussion}
\label{sect:discussion}

 In this paper, we have reported on sensitive X-ray observations performed with \chandra\ and \xmm, which 
 serendipitously  caught \src\ outside outbursts. 
The X-ray luminosity was low and variable in the range from $\sim$$2\times10^{31}$ to 1.1$\times10^{34}$ \lum\ (0.5-10 keV).
%
To summarize this variability, in Fig.\,\ref{fig:uforb} the fluxes corrected for the absorption (0.5-10 keV) versus the orbital phase are displayed.
 Additionally, we analysed  an archival \nustar\ pointing allowing us to investigate the broad-band spectrum  (0.5-30 keV) of \src\ in low state, for the first time.
 A featureless, absorbed power law model is a good deconvolution of the 0.5-30 keV spectrum.

 The absorbing column density is variable within a factor of 10 (as already found and discussed in previous studies; e.g. \citealt{Negueruela2006, Sidoli2009:outburst}),
 with the lowest values ($\sim$10$^{22}$\,cm$^{-2}$) consistent with the optical extinction \citep{Negueruela2006}.
 This variability is usual also in persistent HMXBs and can be ascribed to clumps in the supergiant winds
 or denser wind structures around the compact object \citep{Martinez2017}.

 The light curves observed with \chandra\ and \nustar\
 showed some level of variability, with  flaring activity on a timescale of about $10^{3}$ seconds.
 This behaviour is common to all known SFXTs at any X-ray luminosity (see, e.g., \citealt{Sidoli2019} for a global investigation of \xmm\ observations of SFXTs).
 No statistically significant periodic or quasi-periodic modulation was found in any data set. The limits on their presence, however, are not particularly strict.

 The most remarkable result is the very low X-ray emission caught during the \xmm\ observation, implying  an X-ray luminosity as low as a few $10^{31}$ \lum\ (0.5-10 keV), never observed before in \src.
 A so low X-ray luminosity  has been previously detected in SFXTs only in an another member of the class, the source IGR\,J08408-4503 \citep{Sidoli2021}.
A third case is the 3$\sigma$ upper limit placed on the X-ray luminosity from the SFXT SAX\,J1818.6-1703 at 
$6\times10^{31}$ \lum\ (0.5-10 keV; \citealt{Bozzo2008ul}),
assuming a power law model with $\Gamma$=1.9 and \nh=6$\times$10$^{22}$\,cm$^{-2}$.

 The continuum models used to fit the EPIC spectra of \src\
 were a power law and a thermal emission from a collisionally-ionized plasma ({\tt apec}).
 This latter model is usually adopted to fit the X-ray spectra from early-type massive stars.
 Indeed, the  shocks that naturally develop inside the supersonic strong stellar winds
 heat the wind matter up to X-ray temperatures \citep{Feldmeier1997}.
 The X-ray luminosities from these massive stars  are in the range $10^{31}$-a few $10^{32}$ \lum\
 \citep{Berg1997, Naze2009, Oskinova2016, Nebot2018}.
 This implies that stellar emission from the donor star 
 can contribute a large fraction (or all) of the X-ray luminosity from a SFXT at its lowest state,
 as found recently in the SFXT IGR\,J08408-4503 \citep{Sidoli2021}.

 The adoption of an {\tt apec} model fitting the \xmm\ spectrum
 was also suggested by the resulting steep best-fit power law slope of $\Gamma$=3.5 
 (although we note that the large uncertainty makes this photon index consistent with the values obtained from all other observations reported here).
 The spectral parameters obtained using the {\tt apec} model are consistent with  X-ray stellar emission:
 O-type  stars usually display a double-temperature spectrum, with temperatures
 components $kT_1\approx 0.2$--$0.3$\,keV and $kT_2\approx 0.7$--$0.8$\,keV \citep{Rauw2015, Huenemoerder2020}.
 But the large absorbing column density towards \src\ hampers the detection of the lowest temperature component,
 showing only the hottest one at 0.7 keV.
 The derived luminosity of $6.7\times10^{31}$ \lum\ is also compatible
 with intrinsic X-ray emission from the O-type donor star only.

 We note that the low number of counts in the EPIC spectra and the large uncertainty in the derived spectral parameters, hampers a determination of the X-ray emission contributed by the compact object alone, contrary to what was done in \citet{Sidoli2021} investigating the SFXT IGR\,J08408-4503.
 In that case, the very low absorption allowed us to  clearly disentangle the X-ray emission produced by 
 the supergiant wind from the one contributed by  the compact object.

 However, given the very low level of X-ray luminosity caught from \src, we can apply to \src\ the same physical scenario extensively discussed by \cite{Sidoli2021}, which we briefly summarize here.
In this theoretical picture, a SFXT is thought to be a slowly rotating neutron star (NS) accreting
matter from the supergiant wind \citep{Shakura2012, Shakura2014}.
Because of an inefficient cooling of the accreting matter, a hot shell of matter forms
above the NS magnetosphere and a quasi-spherical settling accretion regime sets-in.
The accretion of the shell material onto the NS occurs by means of the  Rayleigh-Taylor instability
and it is controlled by an inefficient, radiative cooling. This maintains the source in a low X-ray luminosity state,
the most frequent level of emission in SFXTs, at $10^{33}$-$10^{34}$ \lum\ \citep{Sidoli2008}.
In this model, the outbursts in SFXTs can be produced by an instability at the NS magnetosphere
(probably due to magnetic reconnection with the magnetized stellar wind material),
enabling the sudden accretion of the hot shell onto the NS, producing the bright short flares \citep{Shakura2014}. 
\citet{Sidoli2021} explains the very low X-ray luminosity observed
from the SFXT IGR\,J08408-4503  with a NS magnetosphere that is Rayleigh-Taylor stable.
In this regime, the very low X-ray luminosity of $10^{31}$ \lum\ can be produced by the residual accretion onto the NS,
permitted by the mechanism of the Bohm  diffusion through the magnetosphere (see \citealt{Sidoli2021} for a detailed discussion).

It is interesting that  in both sources these lowest luminosity states are shown near the supposed apastron passage.
 However, it is remarkable that the opposite is not true: outbursts in \src\ can happen at any orbital phase, even near apastron \citep{Drave2010}.
  For instance, this is the case of the latest bright flare caught in 2021, that occurred at orbital phase $\phi$=0.97.
The similarity of \src\ with the SFXT IGR\,J08408-4503 also deals with a significant eccentricity of their orbits.
But while in IGR\,J08408-4503 it is well known 
(e=0.63; \citealt{Gamen2015}), in \src\ it is only inferred from the shape of the X-ray orbital light curve  and constrained in the range e=0.16-0.8 \citep{Drave2010}

Finally, it is important to remark that the \xmm\ observation allowed us to increase to $\sim3\times10^{5}$ the source dynamic range, between the X-ray flux at the peak of the flares and in quiescence \citep{Sidoli2009:outburst, Romano2022}.
With this new amplitude of the observed X-ray variability, \src\ can be now classified among the most extremes SFXTs \citep{Sidoli2018}.
This result demonstrates that frequent sensitive observations of transient sources with very low duty cycles
are crucial to better characterize their properties, to catch very  faint  states  and estimate how often they occur.

 \section{Conclusions}
 \label{sect:concl}

We have reported here the analysis of recent \chandra, \xmm\ and \nustar\ observation of the SFXT \src.
All observations caught the source outside outbursts, at a low X-ray luminosity level, below a few 10$^{34}$ \lum.
The main results of our analysis can be summarized as follows:
 \begin{itemize}
 \item the \xmm\ observation has captured the faintest X-ray state ever observed in \src\, at a luminosity of a few 10$^{31}$ \lum, near the supposed apastron passage;
 
 \item this low emission can be fitted equally well by a steep power law model or by a thermal hot plasma emission; this latter is consistent with X-ray emission from the donor star only; however, given the low number of counts, it is difficult to estimate the contribution of the compact object to this very low state;
 
 \item if a fraction of this X-ray emission at a few 10$^{31}$ \lum\ is produced by the compact object, it can be explained by residual accretion onto the NS produced by Bohm diffusion through the NS magnetosphere in the settling accretion regime, similarly to what found in the SFXT  IGR\,J08408-4503 by \citet{Sidoli2021};

 \item the source dynamic range increases to $\sim$3$\times10^5$, making \src\ one of the most extreme SFXTs;
 
 \item the broad-band X-ray spectrum (0.5-30 keV) in the low luminosity state (at 10$^{34}$ \lum) is investigated here for the first time, and it is well fitted by a featureless, absorbed power law model.
 
 \end{itemize}


\begin{acknowledgements}
This work is based on observations performed with \xmm, \chandra\ and \nustar\ satellites. 
\xmm\ is an ESA science mission with instruments and contributions directly funded by ESA Member States and NASA. 
We made use of the software provided by the \chandra\ X-ray Center (CXC) in the application package {\tt CIAO}.
 The \nustar\ mission is a project led by the California Institute of Technology, managed by the Jet Propulsion Laboratory, and funded by the National Aeronautics and Space Administration. This research made use of the $NuSTAR~DAS$ software package, jointly developed by the ASDC (Italy) and Caltech (USA). 
 We made use of the High Energy Astrophysics Science Archive Research Center (HEASARC), a service of the Astrophysics Science Division at NASA/GSFC.
This work has made use of data from the European Space Agency (ESA) mission Gaia (https://www.cosmos.esa.int/gaia), processed by the Gaia Data Processing and Analysis Consortium (DPAC, https://www.cosmos.esa.int/web/gaia/dpac/consortium). Funding for the DPAC has been provided by national institutions, in particular the institutions participating in the Gaia Multilateral Agreement.
LS acknowledges useful information provided by the {\tt xspec} helpdesk.
GP acknowledges funding from the European Research Council (ERC) under the European Union’s Horizon 2020 research and innovation programme (grant agreement No 865637).
We are grateful to the anonymous referee for their prompt and constructive report. 
\end{acknowledgements}

\bibliographystyle{aa}

\end{document}